\documentclass[prl,groupedaddress,twocolumn,floatfix]{revtex4}
\usepackage{amsmath,amsthm,amsfonts,amssymb,graphics,graphicx,epsfig,color,times,natbib,subfigure}

\newcommand{\ket}[1]{\mbox{$ | #1 \rangle $}}

\newcommand{\be}{\begin{equation}}
\newcommand{\ee}{\end{equation}}
\newcommand{\ba}{\begin{eqnarray}}
\newcommand{\ea}{\end{eqnarray}}

\newcommand{\one}{\leavevmode\hbox{\small1\normalsize\kern-.33em1}}

\begin{document}

\title{Characterizing the nonlocal correlations of particles that never interacted}
\author{C. Branciard, N. Gisin, S. Pironio}
\address{Group of Applied Physics, University of Geneva, 20 rue de l'Ecole-de-M\'edecine, CH-1211 Geneva 4, Switzerland}
\date{\today}

\begin{abstract}
Quantum systems that have never interacted can become nonlocally correlated through a process called entanglement swapping. To characterize nonlocality in this context, we introduce local models where quantum systems that are initially uncorrelated are described by uncorrelated local variables. While a pair of maximally entangled qubits prepared in the usual way (i.e., emitted from a common source) requires a visibility close to $70\%$ to violate a Bell inequality, we show that an entangled pair generated through entanglement swapping will already violate a Bell inequality for visibilities as low as $50\%$ under our assumption.
\end{abstract}

\maketitle
It is natural to expect that correlations between distant particles are the result of causal influences originating in their common past --- this is the idea behind Bell's concept of local causality \cite{bell}. Yet, quantum theory predicts that measurements on entangled particles will produce outcome correlations that cannot be reproduced by any theory where each separate outcome is locally determined by variables correlated at the source.  This nonlocal nature of entangled states can be revealed by the violation of Bell inequalities.

However remarkable it is that quantum interactions can establish such nonlocal correlations, it is even more remarkable that particles that never directly interacted can also become nonlocally correlated. This is possible through a process called entanglement swapping~\cite{zukowski_event-ready-detectors_1993}. Starting from two independent pairs of entangled particles, one can measure jointly one particle from each pair, so that the two other particles become entangled, even though they have no common past history. The resulting pair is a genuine entangled pair in every aspect, and can in particular violate Bell inequalities.

Intuitively, it seems that such entanglement swapping experiments exhibit nonlocal effects even stronger than those of usual Bell tests. To make this intuition concrete and to fully grasp the extent of nonlocality in entanglement swapping experiments, it seems appropriate to contrast them with the predictions of local models where systems that are initially uncorrelated are described by uncorrelated local variables. This is the idea that we pursue here.
To precise it further consider the general scenario depicted below.
\begin{figure}[h]
\begin{center}
\includegraphics[width=0.45\textwidth]{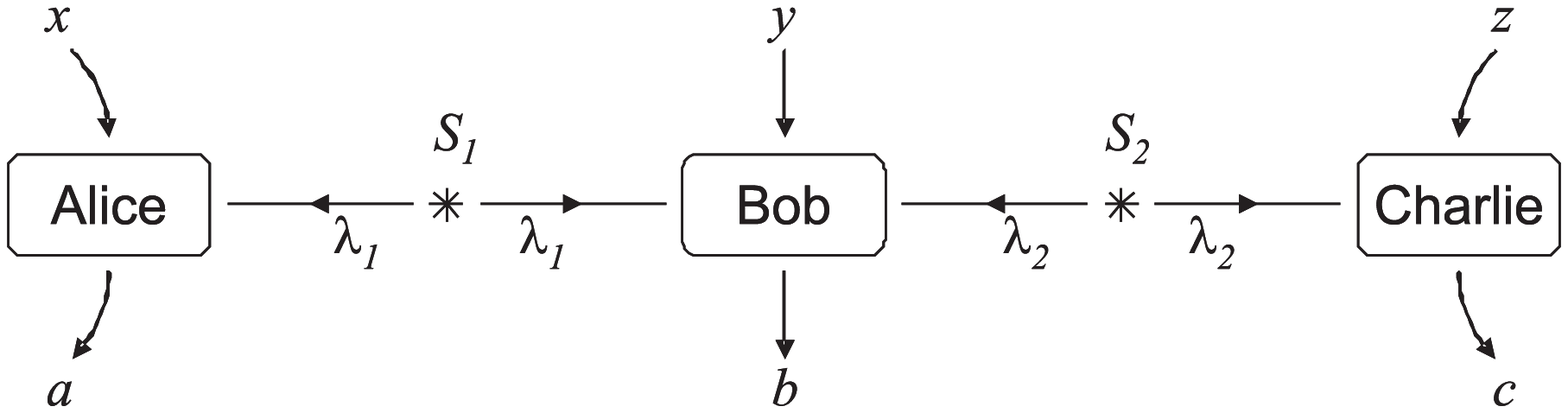}
\end{center}
\end{figure}\\
A source $S_1$ sends particles to Alice and Bob, and a separate source $S_2$ sends particles to Charles and Bob. All parties can perform measurements on their system, labeled $x,y$ and $z$ for Alice, Bob and Charles respectively, and they obtain outcomes denoted $a,b$, and $c$, respectively. Bob's measurement $y$ might correspond to a joint measurement on the two systems that he receives from each source. The correlations between the measurement outcomes of the three parties are described by the joint probability distribution $P(a,b,c|x,y,z)$.
An entanglement swapping experiment is clearly a particular case of this scenario, where Bob's measurement corresponds to a Bell measurement entangling Alice's and Charles's particles.

Under the usual assumption, the tripartite distribution $P(a,b,c|x,y,z)$ would be said to be local if it can be written in the factorized form
\ba && P(a,b,c|x,y,z) \nonumber \\&& \quad= \int \! \mathrm{d} \! \lambda \, \rho(\lambda) \ P(a|x,\lambda)P(b|y,\lambda)P(c|z,\lambda) \,, \label{eq_locality1} \ea
where the variable $\lambda$ with distribution $\rho(\lambda)$ describes the joint state of the three systems according to the local model, and  $P(a|x,\lambda)$, $P(b|y,\lambda)$, $P(c|z,\lambda)$ are the local probabilities for each separate outcome given $\lambda$.

In our scenario, however, they are two separate sources $S_1$ and $S_2$. It it thus natural to assume that the local model assigns two different states $\lambda_1$ and $\lambda_2$, one to each source, and to consider instead of (\ref{eq_locality1}) the decomposition
\ba \label{eq_locality2}&& P(a,b,c|x,y,z) \\  && \quad = \int \!\!\!\!\! \int \! \mathrm{d} \! \lambda_1 \mathrm{d} \! \lambda_2 \, \rho(\lambda_1,\lambda_2) \ P(a|x,\lambda_1)P(b|y,\lambda_1,\lambda_2)P(c|z,\lambda_2) . \nonumber   \ea
The local response function for Alice now depends only on $\lambda_1$, the one of Charles only on $\lambda_2$, while the one of Bob depends on both $\lambda_1$ and $\lambda_2$. So far, the decompositions (\ref{eq_locality1}) and (\ref{eq_locality2}) are equivalent because  $\rho(\lambda_1,\lambda_2)$ could be different from zero only when $\lambda_1=\lambda_2=\lambda$ to recover (\ref{eq_locality1}). We now introduce our basic assumption: since the two sources $S_1$ and $S_2$ are supposed to be independent and uncorrelated, it is natural to assume that this property carries over to the local model. The variables $\lambda_1$ and $\lambda_2$ should therefore be independent and their joint distribution $\rho(\lambda_1,\lambda_2)$ factorize:
\ba \rho(\lambda_1,\lambda_2) = \rho_1(\lambda_1)\rho_2(\lambda_2) . \label{eq_indep_loc_constr} \ea
We refer to models satisfying this independence assumption as ``bilocal" models, since they aim at explaining the correlations $P(a,b,c|x,y,z)$ with two independent sources of local variables.

Even though the local variables $\lambda_1$ and $\lambda_2$ are initially independent, once conditioned on the joint measurement result of Bob they will bear enough correlations to reproduce non-trivial correlations between Alice's and Charles's system. These correlations, however, are much weaker than those that can be established through joint measurements in quantum theory. We introduce below a (quadratic) Bell inequality that is satisfied by all bilocal correlations, but which is violated by quantum correlations. As we will see, our inequality simplifies the requirements for the demonstration of quantumness in entanglement swapping experiments.

Restricted classes of local models with independent sources were considered in \citep{gisin-gisin,GHZ} within the context of the detection loophole. But apart from these exploratory works, little was known about how nonlocality is induced through measurements on independent quantum systems. Beyond its fundamental interest, nonlocality is also known to play a key role in several quantum information protocols \cite{cc,DI}, and measurement-induced correlations are at the basis of quantum repeaters \cite{qr} and measurement-based quantum computation \cite{mbqc}. One of our contribution is to introduce a theoretical framework to address broadly the role of nonlocality in such contexts.

Before entering in the details of our results, it might be worth justifying further our independence assumption. It is strictly speaking an assumption, rather than something which follows logically from locality. Indeed, some events in the common past of the sources $S_1$ and $S_2$ could in principle have influenced, in a way that is perfectly in accord with locality, both $\lambda_1$ and $\lambda_2$ such that they wind up correlated, in violation of (\ref{eq_indep_loc_constr}). However, an assumption similar to (\ref{eq_indep_loc_constr}) is actually hidden in any standard Bell-type experiment. In order to derive a Bell-type inequality, one needs (in addition to the premise of local causality) an assumption having to do with the measurement settings being ``freely chosen". What this means in practice is that the measurement settings are determined by a random mechanism that is considered independent of the variable $\lambda$ describing the particle source. Seen from this perspective, the assumption that the laser sources in the quantum random number generators used to choose the measurement settings in a standard Bell experiment~\cite{weihs} are independent of the laser source generating the entangled photons is not much different from the assumption that the two laser sources (which may be of different brands, assembled in different parts of the world, and powered by different electrical supplies) used in an entanglement swapping experiment are independent. Of course we cannot exclude in principle that such apparently independent sources are significantly correlated. But, quoting Bell, ``this way of arranging quantum mechanical correlations would be even more mind-boggling than one in which causal chains go faster than light. Apparently separate parts of the world would be deeply and conspiratorially entangled" \cite{bell}.

\paragraph{Characterization of the bilocal set.}
We start by giving a characterization of the set of bilocal correlations which is more handy for analytical and numerical purposes than the definition (\ref{eq_locality2}) and (\ref{eq_indep_loc_constr}). First note that without loss of generality the local response function $P(a|x,\lambda_1)$ of Alice can be taken to be deterministic, i.e., such that it assigns a unique measurement output $a$ to every input $x$ (any randomness used locally by Alice can always be thought of as being included in the shared variable $\lambda_1$). In the case of a finite number of possible measurement inputs and outputs, there is a finite number of such deterministic strategies corresponding to an assignment of an output $\alpha_x$ to each of Alice's $N$ possible inputs $x$. We thus label each of these strategies with the string $\alpha=\alpha_1\ldots \alpha_N$ and denote the corresponding response function $P_\alpha(a|x)$. Similarly, the response functions $P(b|y,\lambda_1,\lambda_2)$ and $P(c|z,\lambda_2)$ can also be taken deterministic. We label the associated strategies $\beta$ and $\gamma$ and the corresponding response functions $P_\beta(b|y)$ and $P_\gamma(c|z)$.

Let $\Lambda^{12}_{\alpha\beta\gamma}$ denote the set of pairs $(\lambda_1,\lambda_2)$ specifying the strategies $\alpha$, $\beta$, and $\gamma$ for Alice, Bob, and Charles. Defining $q_{\alpha\beta\gamma}=\int\!\!\!\int_{\Lambda^{12}_{\alpha\beta\gamma}} \mathrm{d} \! \lambda_1 \mathrm{d} \! \lambda_2 \, \rho(\lambda_1,\lambda_2)$, Eq. (\ref{eq_locality2}) can then be rewritten as
\be P(a,b,c|x,y,z) = \sum_{\alpha,\beta,\gamma} q_{\alpha\beta\gamma} \, P_{\alpha}(a|x)P_{\beta}(b|y)P_{\gamma}(c|z)
\label{eq_locality_sum} \ee
with $q_{\alpha\beta\gamma}\geq 0$ and $\sum_{\alpha\beta\gamma} q_{\alpha\beta\gamma}=1$.
So far we have not used the independence condition (\ref{eq_indep_loc_constr}), and (\ref{eq_locality_sum}) corresponds to the well-known decomposition of local correlations as a convex sum of deterministic strategies, where the weights $q_{\alpha\beta\gamma}$ can be understood as the probabilities assigned by the source to the strategies $\alpha$, $\beta$, and $\gamma$.

Let us now define $q_{\alpha\gamma}=\sum_\beta q_{\alpha\beta\gamma}$, $q_{\alpha}=\sum_{\beta\gamma} q_{\alpha\beta\gamma}$, and $q_{\gamma}=\sum_{\alpha\beta} q_{\alpha\beta\gamma}$. Using the fact that $\Lambda^{12}_{\alpha,\beta\,\gamma}=\left(\Lambda^1_\alpha\times\Lambda^2_\gamma\right)\cap \Lambda^{12}_\beta$, as follows from (\ref{eq_locality2}), the independence condition (\ref{eq_indep_loc_constr}) implies that
\be q_{\alpha\gamma} = q_{\alpha} q_{\gamma}\,. \label{eq_biloc_constr}
\ee
Conversely, any correlation $P(a,b,c|x,y,z)$ satisfying (\ref{eq_locality_sum}) and (\ref{eq_biloc_constr}) can be written in the form (\ref{eq_locality2}). Indeed, since $q_{\alpha\gamma}=q_\alpha q_\gamma$, we can write $q_{\alpha\beta\gamma}=q_\alpha q_\gamma q_{\beta|\alpha\gamma}$. Inserting this expression in (\ref{eq_locality_sum}) and defining $P_{\alpha,\gamma}(b|y)=\sum_\beta q_{\beta|\alpha\gamma} P_\beta(b|y)$, we then find that $P(a,b,c|x,y,z) = \sum_{\alpha,\gamma} q_\alpha q_\gamma \, P_{\alpha}(a|x)P_{\alpha,\gamma}(b|y)P_{\gamma}(c|z)$, which is clearly of the form (\ref{eq_locality2}). We thus conclude that a tripartite correlation is bilocal if and only if it admits the decomposition (\ref{eq_locality_sum}) with the restriction (\ref{eq_biloc_constr}).

The bilocal set that we have just characterized is clearly contained in the local set. The extremal points of the local set, corresponding to deterministic strategies, are also bilocal, but a mixture of deterministic strategies is not necessarily bilocal due to the non-convex constraint (\ref{eq_biloc_constr}). Therefore, one cannot use standard Bell inequalities to distinguish one set from the other. As we will see below, however, correlations can be shown to be non-bilocal using \emph{non-linear} inequalities (or joint sets of linear inequalities).
Note that while deciding if a given set of correlations is local can be conveniently solved using linear programming, deciding if a correlation is bilocal is a quadratically constrained problem which is much more difficult to handle numerically. However, standard linear and semidefinite relaxations can be used to deal with the non-linear constraint (\ref{eq_biloc_constr}). We describe in \cite{prep}, a linear relaxation of (\ref{eq_biloc_constr}) which works well on many instances.

\paragraph{Application to entanglement swapping.}
We now illustrate how the bilocality constraint restricts the set of possible correlations on a simple example inspired by the standard entanglement swapping protocol. The sources $S_1$ and $S_2$ send pairs of particles in the singlet state $\ket{\Psi^-}=\left(|01\rangle-|10\rangle\right)/\sqrt{2}$. Bob performs a Bell state measurement on the two particle he receives, with four possible outputs $b=b_0b_1=00,01,10,11$ corresponding to the four Bell states $|\Phi^+\rangle$, $|\Psi^+\rangle$, $|\Psi^-\rangle$, and $|\Phi^-\rangle$, respectively. Depending on Bob's result, Alice and Charles's particles end up in the corresponding Bell state. To check whether the entanglement swapping succeeded, we assume that Alice and Charles can perform one out of two measurement $x,z\in\{0,1\}$ with binary outputs $a,c\in\{0,1\}$ on their system. This is sufficient, e.g., to test the CHSH inequality~\cite{chsh} (or more precisely, for each state prepared by Bob, a different version of the CHSH inequality corresponding to a relabeling of the inputs and outputs). This scenario is characterized by the probabilities $P_Q(abc|xz)=P_Q(b)P_{Q|b}(ac|xz)$, where $P_{Q|b}(ac|xz)$ denote the correlations seen by Alice and Charles conditioned on Bob's output $b$, and where for convenience we omitted Bob's input $y$ since he is assumed to make a single, fixed measurement.

We are interested in the robustness of $P_Q$ to the admixture of white noise, quantified by the maximal $v\in[0,1]$ such that $P_Q(v)=vP_Q+(1-v)P_R$ is bilocal, where $P_R$ denotes the distribution with completely random outcomes. The quantity $v$ can also be interpreted as the experimental visibility of the final entangled pair shared by Alice and Charles.

If Alice and Charles use the measurement settings optimal for the CHSH inequality, given by $x_0=\sigma_x$, $x_1=\sigma_z$, $z_0=\left(\sigma_x+\sigma_z\right)/\sqrt{2}$, and $z_1=\left(\sigma_x-\sigma_z\right)/\sqrt{2}$, no improvement is obtained over the usual locality condition, i.e., we found that the quantum correlations become bilocal for a visibility $v=1\sqrt{2}$, the same point at which they also become local. Using the characterization defined by (\ref{eq_locality_sum}) and (\ref{eq_biloc_constr}), we looked numerically for other choices of measurement settings for Alice and Charles and the best noise resistance that we found is $v=1/2$ and is obtained for $x_0=z_0=(\sigma_x+\sigma_z)/\sqrt{2}$, $x_1=z_1=(\sigma_x-\sigma_z)/\sqrt{2}$. The corresponding correlations observed by the three parties are given by $P_Q(b)=1/4$ for all $b$, and
\be P_{Q|b}(ac|xz) = \left\{ \begin{array}{cl} 1/2 & \mathrm{if \ } a \oplus c=b_0 \mathrm{\ and \ } x \oplus z=b_1 \\
0 & \mathrm{if \ } a \oplus c\neq b_0 \mathrm{\ and \ } x \oplus z =b_1 \\
1/4 & \mathrm{otherwise}\end{array} \right.\label{eq_decomp_PQ}
\ee
For instance, if Alice and Charles end up in the $\ket{\Phi^+}$ state, corresponding to $b=00$, they obtain perfectly correlated results if they performed the same measurements, and completely uncorrelated results otherwise.
The above correlations are local as they can be decomposed as $P_Q=(P_C+\bar P_C)/2$ where $P_C$ and $\bar P_C$ are defined in term of deterministic strategies as $P_C=\frac{1}{8}\sum_{\alpha \beta_0\beta_1}P_{\alpha\alpha}P_{\beta_0\beta_1}P_{(\alpha\oplus \beta_0)(\alpha\oplus \beta_0)}$ and $\bar P_C=\frac{1}{8}\sum_{\alpha \beta_0\beta_1}P_{(\alpha\oplus \beta_1)(\alpha\oplus \beta_1\oplus 1)}P_{\beta_0\beta_1}P_{(\alpha\oplus \beta_0)(\alpha\oplus \beta_0\oplus+1)}$. They are not bilocal, however, as we now show.

\paragraph{A quadratic Bell inequality for bilocality.}
Let us first define, for a given probability distribution $P(abc|xz)$, the following correlation term between Alice and Charles' outputs, conditioned on Bob's output:
\be E_b(xz)=\sum_{a\oplus c=b_0} P_b(ac|xz) - \sum_{a\oplus c\neq b_0} P_b(ac|xz)\,. \ee
Inspired by the properties (\ref{eq_decomp_PQ}) of $P_Q$, we introduce the following combination that quantifies the high degree of correlations expected between Alice's and Charles's outcomes when $x\oplus z=b_1$
\be I=\sum_b P(b) \sum_{x\oplus z=b_1} E_b(xz)\,. \ee
We quantify the deviation from the expected uncorrelated results when $x\oplus z\neq b_1$ through
\be E=\max_{b}\max_{x\oplus z\neq b_1} 4|P(b)E_b(xz)| \,.\ee
For $P_Q$, one gets $I=2$ and $E= 0$.  Note that there exist bilocal correlations which also reach the value $I= 2$, for instance, the deterministic point defined by $\alpha=00, \beta=00, \gamma=00$, or the correlations $P_C$ introduced above. For these correlations, $E = 4$ and $E=1$, respectively. The linear expression $I$ cannot therefore be used as a standard Bell inequality to test bilocality. However, as we prove below, the following  quadratic inequality
\be I\leq 1+E^2 \label{quadrineq}\ee
is satisfied by all bilocal correlations and is violated by the quantum point $P_Q$, since we find $2>1+0$. Intuitively, $I$ correspond to a Bell inequality whose bound is not fixed for the entire bilocal set, but depends on how much the outputs are uncorrelated when $x\oplus z\neq b_1$, as quantified by $E$.

\begin{figure}
\begin{center}
\includegraphics[width=0.3\textwidth]{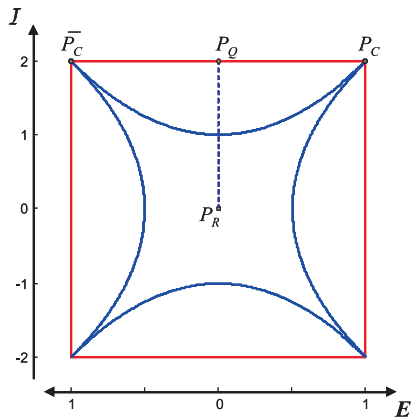}
\end{center}
\caption{Two-dimensional slice of the correlation space, which contains the points $P_Q, P_C, \bar{P_C}$ and $P_R$ defined in the text [this is precisely the slice that contains all the depolarized correlations $P''$ introduced in the proof of (\ref{quadrineq}); they can indeed be written as $P'' = X P_C + Y \bar{P_C} + (1-X-Y)P_R$]. The square delimits the local polytope in this slice. All these local correlations can be reproduced in quantum theory with two independent sources. They cannot, however, be all reproduced locally with two independent sources. The four portions of parabola delimit the bilocal set [the upper parabola is obtained from (\ref{quadrineq}). A similar constrained Bell-type inequality can be derived to obtain the lower, left, and right parabolas]. The quantum point $P_Q$ enters the bilocal region for a visibility $v\leq 1/2$.
\label{fig_slice}}
\end{figure}

The noisy point $P_Q(v)$ yields $I(v)=2v$ and $E(v)=0$ and thus violate (\ref{quadrineq}) whenever $v>1/2$ (see Fig.~\ref{fig_slice}). On the other hand, $P_Q(v)$ is bilocal when $v\leq 1/2$~\footnote{$P_Q(v)$ can be decomposed, using the notation of the proof of (\ref{quadrineq}), as $P''$ with $r=s=1/2$, $t=u=1/2+v$.}. Our inequality thus detects optimally the resistance to noise of the point $P_Q$.
\paragraph{Proof of (\ref{quadrineq}).}
Let $P$ be a bilocal probability distribution with decomposition (\ref{eq_locality_sum}), where $\alpha=\alpha_0\alpha_1$, $\beta=\beta_0\beta_1$, $\gamma=\gamma_0\gamma_1$. Let us define the following weights (where $\bar\alpha_0 = \alpha_0 \oplus 1$ and similarly for the other indices):
\ba q_{\alpha_0\alpha_1,\beta_0\beta_1,\gamma_0\gamma_1}' = (q_{\alpha_0\alpha_1,\beta_0\beta_1,\gamma_0\gamma_1} + q_{\alpha_0\alpha_1,\bar\beta_0\beta_1,\bar\gamma_0\bar\gamma_1} \qquad \nonumber \\ + q_{\bar\alpha_0\bar\alpha_1,\bar\beta_0\beta_1,\gamma_0\gamma_1} + q_{\bar\alpha_0\bar\alpha_1,\beta_0\beta_1,\bar\gamma_0\bar\gamma_1})/4 , \label{eq_depol1} \nonumber\\ q_{\alpha_0\alpha_1,\beta_0\beta_1,\gamma_0\gamma_1}'' = (q_{\alpha_0\alpha_1,\beta_0\beta_1,\gamma_0\gamma_1}' + q_{\alpha_0\alpha_1,\beta_0\bar\beta_1,\gamma_1\gamma_0}' \qquad \nonumber \\ + q_{\alpha_1\alpha_0,\beta_0\bar\beta_1,\gamma_0\gamma_1}' + q_{\alpha_1\alpha_0,\beta_0\beta_1,\gamma_1\gamma_0}')/4. \label{eq_depol2}\nonumber \ea
The ``depolarized" correlation $P''=\sum_{\alpha\beta\gamma} q_{\alpha\beta\gamma}'' P_\alpha P_\beta P_\gamma$ is then also bilocal, i.e., $q_{\alpha\gamma}''=q_\alpha''q_\gamma''$. Moreover, $P''$ is such that $I''=I$ and $E''\leq E$.  Due to the symmetries imposed through the above equations, the weights $q''_{\alpha\beta\gamma}$ depend on only 4 parameters, which we choose to be $r=q''_{\alpha_0=\alpha_1}, s=q''_{\gamma_0=\gamma_1}, t=q''_{\beta_0=0|\alpha_0\alpha_1=00,\gamma_0\gamma_1=00}$ and $u=q''_{\beta_0=\beta_1|\alpha_0\alpha_1=01,\gamma_0\gamma_1=01}$ (with obvious notations, the weights $q''_{\alpha\beta\gamma}$ being understood as probabilities~\footnote{For instance, $r=q''_{\alpha_0=\alpha_1}=\sum_{\beta\gamma} (q''_{00,\beta,\gamma}+q''_{11,\beta,\gamma})$. Note that $r,s,t,u \in [0,1]$. The weights $q''_{\alpha\beta\gamma}$ are then fully defined by $q''_{\alpha_0\alpha_1}=\frac{r}{2} \delta_{\alpha_0,\alpha_1} + \frac{1-r}{2} \delta_{\alpha_0,\bar \alpha_1}$; $q''_{\gamma_0\gamma_1}=\frac{s}{2} \delta_{\gamma_0,\gamma_1} + \frac{1-s}{2} \delta_{\gamma_0,\bar \gamma_1}$; $q''_{\beta|\alpha\gamma} = \frac{t}{2} \delta_{\alpha_0 \oplus \gamma_0,\beta_0} + \frac{1-t}{2} \delta_{\alpha_0 \oplus \gamma_0,\bar\beta_0}$ if $\alpha_0=\alpha_1$ and $\gamma_0=\gamma_1$; $q''_{\beta|\alpha\gamma} = \frac{u}{2} \delta_{\alpha_0 \oplus \gamma_0,\beta_0\oplus\beta_1} + \frac{1-u}{2} \delta_{\alpha_0 \oplus \gamma_0,\beta_0\oplus\beta_1\oplus 1}$ if $\alpha_0\neq \alpha_1$ and $\gamma_0\neq \gamma_1$; $q''_{\beta|\alpha\gamma} = \frac{1}{4}$ otherwise.}). Defining $X = rs(2t-1)$ and $Y = (1-r)(1-s)(2u-1)$, we find
$I'' = 2(X+Y)$, $E''= |X-Y|$. From their definition, $X$ and $Y$ are restricted to satisfy $\sqrt{|X|}+\sqrt{|Y|} \leq \sqrt{rs}+\sqrt{(1-r)(1-s)} \leq 1$. One can easily check that under this constraint, $I'' \leq 1+E''^2$, which implies (\ref{quadrineq}).

Note that for any value of $E \leq 1$, the bound is tight, i.e., there exists a bilocal correlation such that $I=1+E^2$. Take for instance $P''$, with $r=s=(1+E)/2$ and $t=u=1$.\hfill$\Box$

\paragraph{Discussion and open questions.}
We have shown that if one makes the reasonable assumption, underlying all of modern empirical science, that the world is composed of different parts that are independent for the purposes at hand, then nonlocality is a phenomena even more common than usually thought. While the standard analysis leads to the conclusion that the final singlet pair in an entanglement swapping experiment needs a visibility higher than $v=1/\sqrt{2}\simeq 71\%$ to violate the CHSH inequality and will not violate any Bell inequality (with von-Neumann measurements) for visibilites smaller than $v\simeq 66\%$ \cite{grod}, we have shown here that under our assumption it already exhibits nonlocal correlations for visibilites as low as $50\%$. This simplifies the requirements for the demonstration of quantumness in entanglement swapping experiments \cite{matheus}.

Is this $v=50\%$ limit a fundamental limit? It is easy to show that there exists a bilocal model for visibilites lower than $25\%$ \cite{prep}. But what happens in-between? Can we lower the visibility threshold by  considering experiments with more inputs? Do we gain by letting Bob choose between two or more measurements, e.g., between two Bell state measurements in different bases? Can bilocality be violated for visibilites lower than $33\%$, corresponding to a final noisy singlet pair that is separable? This last question is not completely trivial at first sight: a setting where the source $S_1$ produces a singlet state, $S_2$ the \emph{separable} state $\rho=(|{+}z,{+}z\rangle\langle{+}z,{+}z|+|{+}x,{-}z\rangle\langle{+}x,{-}z|)/2$, where Bob performs a standard Bell measurement, Alice measures in the $(\sigma_x\pm\sigma_z)/\sqrt{2}$ bases and Charles always measures in the $z$ basis generates correlations that are non-bilocal~\footnote{This experiment can be interpreted as a standard test of the CHSH inequality performed on the singlet $S_1$, where Charles's outcomes specify in which basis ($x$ or $z$) Bob's particles has been measured. Such an interpretation makes the similarity between our independence assumption and the ``free choice" assumption of standard Bell experiments more explicit~\cite{prep}.}. This shows that a Bell measurement can correlate independent systems in ways that are even more astonishing that one would ``quantum naively" think. From this perspective, it would be interesting to characterize the class of  states (including separable states) that are non-bilocal when correlated through Bell measurements.

The present work raise many other questions.  In particular, the condition (\ref{eq_indep_loc_constr}) can be straightforwardly extended to models with $n$ independent sources. How do such $n$-local models differ from bilocal ones? How does their tolerance to noise scale with $n$? Finally, it would be interesting to explore the implications of our approach and findings in the context of quantum information protocols based on non-locality, and in particular protocols that use at their heart measurements on independent systems, such as quantum repeaters and measurement based computation.

We acknowledge support by the Swiss NCCR \textit{Quantum Photonics} and the European ERC-AG \textit{QORE}.

\bibliography{bib_bilocality}

\end{document}